\documentstyle[aps,preprint]{revtex}
\begin{document}
\author{\"Ors Legeza}
\title{Stability of the Haldane phase in anisotropic magnetic ladders} 
\address{Research Institute for Solid State Physics,
       H-1525 Budapest, P.\ O.\ Box 49, Hungary}
\address{Technical University of Budapest,
       H-1521 Budapest, Hungary}
\author{Jen\H{o} S\'olyom}
\address{Research Institute for Solid State Physics,
       H-1525 Budapest, P.\ O.\ Box 49, Hungary}

\date{\today}
\maketitle
\begin{abstract}
We have considered the properties of anisotropic two-leg ladder models
with $S=1/2$ or $S=1$ spins on the rungs, using White's density matrix 
renormalization group method. We have generalized the method by taking 
into account the symmetries of the model in order to reduce the 
dimensions of the matrix to be diagonalized, thereby making possible to 
consider more states. The boundaries in the parameter space of the 
extended region, where the Haldane phase exists, are estimated.

\end{abstract}
\bigskip
\narrowtext
\section{Introduction}

The properties of magnetic systems where the localized moments form a ladder
like structure \cite{johnston,hiroi,cava}, have been intensively 
studied recently. Most of such materials can be described by an 
isotropic spin Hamiltonian, therefore the theoretical studies were almost 
exclusively done on isotropic models 
\cite{hida,dagotto,watanabe,okamoto,hsu,noack,gopalan}. 
The question arises as to what effect the anisotropy might have. The 
answer to the question is not trivial.

It is known that two-leg ladder models with $S=1/2$ spins on both legs
behave basically like an $S=1$ spin model if the rung coupling between the
spins on the two legs is strong enough. It is also known that integer and 
half-odd integer spin models have essentially different phase diagrams,
when considered as a function of the anisotropy of the exchange.
Beside the ferromagnetic, antiferromagnetic and planar phases a new one, 
the Haldane phase \cite{haldane} appears for a finite range of anisotropy. 
Our aim in this paper is to study, how this new phase appears in 
anisotropic ladder models for intermediate values of the interchain
couplings.  

For this purpose we have determined the low lying part of the energy
spectrum of an anisotropic Heisenberg ladder model by using the density matrix
renormalization group (DMRG) method proposed recently by White \cite{white}.
We have generalized the method in such a way that together with any spin 
configuration that is kept after the truncation of the Hilbert space, the 
other configurations related to it by the symmetries of the ladder model 
are also automatically taken into account. This allows to consider more 
states without increasing the size of the matrices and thereby to improve 
the accuracy.

The layout of the paper is as follows. In Sec. II we give a
short description of the ladder model and show what phase diagrams are
expected, if the value of the spins on the legs is $S=1/2$ or $S=1$. The 
symmetry considerations introduced in the application of the DMRG procedure 
are briefly discussed in Sec III. Sec. IV presents the numerical results. 
Finally Sec V. contains a brief discussion of the results.

\section{Ladder models and their expected phase diagram}

In a two-leg ladder model the spins at rung $i$ will be denoted
by $\sigma_{i}$ and $\tau_{i}$. The length of the spin can be 
arbitrary. In most of the calculations we will take $S=1/2$ spins, 
but a comparison with the $S=1$ case will also be considered.

When the coupling between the legs is neglected, the Hamiltonian of
two decoupled anisotropic Heisenberg chains is recovered. It can be 
written in the form,
\begin{eqnarray}
{\cal H}_0 & = &\sum_i\left[ \frac{1}{2} J_{xy} (\sigma_i^+\sigma_{i+1}^- +
\sigma_i^-\sigma_{i+1}^+ )+ J_{z}\sigma_i^z\sigma_{i+1}^z\right]\nonumber\\
    & & + \sum_i\left[ \frac{1}{2} J_{xy} (\tau_i^+\tau_{i+1}^- +
\tau_i^-\tau_{i+1}^+ )+ J_{z}\tau_i^z\tau_{i+1}^z\right]\/.
\label{eq:h0}
\end{eqnarray}
The $xy$ part of the exchange is written in terms of the raising and lowering
operators, and its coupling is allowed to differ from that of the 
$zz$ part. 

Introducing now the couplings between the chains, different kinds of 
ladder models can be constructed depending of the choice of this 
coupling. Usually this coupling is assumed to act between spins on the 
same rung only. In an anisotropic model this would mean a coupling of the form,
\begin{equation}
    {\cal H}_1  =  \sum_i\left[ \frac{1}{2} J'_{xy} (\sigma_i^+\tau_{i}^- +
    \sigma_i^-\tau_{i}^+ )+ J'_{z}\sigma_i^z\tau_{i}^z\right]  \/.
\end{equation}
In the limit when this coupling is ferromagnetic and strong, the two
spins on the same rung couple into a single larger spin. That is the 
reason why a spin-1/2 ladder can behave like a spin-1 chain, or a 
spin-1 ladder like a spin-2 chain.  

Assuming that the diagonally situated nearest neighbour spins are also 
coupled, it was shown\cite{white96} that this coupling makes the
hidden topological long-range order characteristic for the Haldane
phase even stronger. 

An alternative way to introduce interchain coupling is to choose the 
following form for the interaction,
\begin{eqnarray}
  {\cal H}_1 & = & \sum_i\left[ \frac{1}{2} J'_{xy} (\sigma_i^+\tau_{i+1}^- +
  \sigma_i^-\tau_{i+1}^+ )+ J'_{z}\sigma_i^z\tau_{i+1}^z\right]\nonumber\\
  & & +  \sum_i\left[ \frac{1}{2} J'_{xy}(\tau_i^+\sigma_{i+1}^- +
  \tau_i^-\sigma_{i+1}^+ )+ J'_{z}\tau_i^z\sigma_{i+1}^z\right]\/.
\label{eq:h1}
\end{eqnarray}
In this paper we look at the properties of the model defined by Eqs.\
(\ref{eq:h0}) and (\ref{eq:h1}). 
In a special case this
is equivalent to one of the models studied by White\cite{white96},
but we will consider a larger parameter space, although 
the an\-iso\-tropy ratio will be 
taken to be the same for the intrachain and interchain couplings,
\begin{equation}
       J_{z} / J_{xy} =  J'_{z} / J'_{xy} \/,
\end{equation}
and therefore a single parameter 
\begin{equation}
   \lambda = { J'_{xy} \over J_{xy} } = { J'_{z} \over J_{z} }
\label{eq:lambda}
\end{equation}
will be used to characterize the strength of the interchain couplings.

An interesting feature of our model is that by interchanging the spins 
between the two legs on every even or odd site, the intrachain and interchain
couplings change their role. At $\lambda =1$, where the intrachain and 
interchain couplings are equal, the model is transformed into itself. 
For other values of $\lambda$, the energy scale has to be changed. Taking 
this into account, we find that for $\lambda > 0$ all energy levels 
satisfy a simple self-duality relationship\cite{solyom1},
\begin{equation}
     E(\lambda )=\lambda E(1/\lambda )\/.
\label{eq:self}
\end{equation}
This allows to connect the weak- and strong-coupling limits of the model. 
For $\lambda <0$  the model would include 
ferromagnetic and antiferromagnetic interchain couplings, which 
requires the modification\cite{legeza} of Eq.\ (\ref{eq:self}). 

The self-dual $\lambda=1$ point plays a special role in other respects 
as well. At this point, with our choice of the couplings, the Hamiltonian 
of a single chain with composite spin is recovered,
\begin{equation}
   {\cal H}= \sum_i \left[ J_{xy}\left( S_i^x S_{i+1}^x +
    S_i^y S_{i+1}^y\right) + J_z S_i^z S_{i+1}^z \right] \/,
\label{eq:heis}                             
\end{equation}
where $\vec S_i=\vec\sigma_i +\vec\tau_i$. It is obvious, that
if both $\sigma$ and $\tau$ are spin-1/2 operators, this model is not a 
true $S=1$ Heisenberg model, since the two spins can form not only the 
triplet, but also the singlet combination. It has been shown\cite{solyom1}, 
however, that the low lying levels of the Hamiltonian in Eq.\ (\ref{eq:heis}) 
and that of the true $S=1$ Heisenberg model are identical, and therefore they 
have the same phase diagram. 

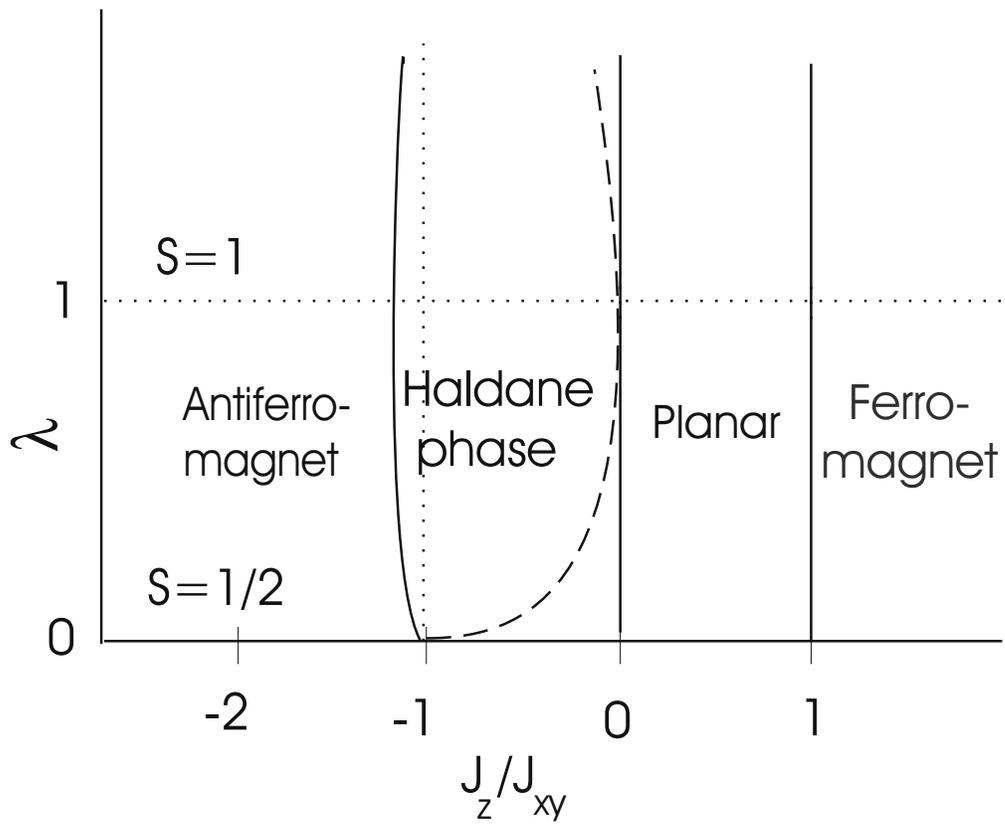
\begin{figure}
\caption{Schematic phase diagram of the $S=1/2$ ladder model in the
$(\lambda, J_z/J_{xy})$ plane. The dashed line shows an alternative phase
boundary between the Haldane and planar phases, as discussed in the text.}
\label{fig:phase}
\end{figure}

The phase diagram expected for the model in the plane spanned by the 
anisotropy, $J_z/J_{xy}$ and $\lambda$, is shown schematically in Fig.\ 
\ref{fig:phase}. In drawing the phase boundaries we have taken into 
account that the ferromagnetic phase appears for $J_{z}/ J_{xy} \ge 1 $, 
independently of $\lambda$. On the other hand the boundary between the 
Haldane and antiferromagnetic phases is expected to depend on $\lambda$.  
At $\lambda =0$ the ladder model consists of two decoupled spin-1/2 
chains, and therefore the Haldane phase does not exist. At $\lambda = 1$, 
according to previous finite-size 
cal\-cu\-la\-tions\cite{botet,gomez,nomura,sakai} on the equivalent
$S=1$ chain, the massive Haldane phase exists in a finite range of 
anisotropy. The numbers obtained for the critical values of the 
anisotropy are somewhat different in the different calculations, but 
the best estimates give the range 
\begin{equation}
     -1.18 < J_z/J_{xy} < 0 \/.
\end{equation}
The value of the critical anisotropy, $(J_z/J_{xy})_{c1} \simeq -1.18$, where 
the transition from the Haldane phase with singlet ground state to the
doubly degenerate antiferromagnetic state occurs, is not determined by 
any symmetry. It is an Ising type transition. Close to it, on both sides 
of the transition point the gap opens linearly. 

It follows from the duality relationship for the energy levels that if 
$\lambda_c$ is a critical point, then $1 / \lambda_c$ should also be a 
critical coupling. The phase boundary of the 
antiferromagnetic phase should therefore start from $(J_{z}/ J_{xy})_{c1} 
= - 1 $ at $\lambda=0$, returning to this value when $\lambda \rightarrow 
\infty$, and passing through $(J_z/J_{xy})_{c1} \simeq -1.18$ at 
$ \lambda = 1$.

Looking at the boundary between the Haldane and planar phases, the 
massive phase appears from the critical planar phase via a 
Kosterlitz-Thouless type transition. In the continuum limit of the $S=1$
chain the value $(J_{z}/ J_{xy})_{c2} = 0$ is obtained \cite{nijs} for the 
critical anisotropy. This particular value suggests that it is determined 
by a hidden symmetry of the model, and therefore it is expected that 
this same critical value could be found in the $S=1/2$ spin ladder model 
at any $\lambda$. This would mean that for $ -1 \leq J_{z}/ J_{xy} < 0$ the 
interchain coupling is a relevant perturbation. An arbitrarily small 
coupling will already generate a Haldane gap. The corresponding phase 
boundary is shown in  Fig.\ \ref{fig:phase} by a vertical straight 
line at $J_z/J_{xy}=0$.

This natural assumption becomes questionable, however, if we compare
this phase diagram with that of the spin-1 ladder. Based on similar 
considerations we can argue that at $\lambda=0$ the behavior is that of 
two decoupled spin-1 chains, while at $\lambda=1$ the properties of the 
spin-2 Heisenberg model should be recovered. Since both are integer 
spin models, the Haldane phase exists in both cases in an extended 
range of anisotropy around the isotropic antiferromagnetic point. In the 
spin-2 model this range is, however, much narrower\cite{alcaraz} than 
in the spin-1 case, so clearly the boundaries of the Haldane phase 
to both the planar and the antiferromagnetic phases should strongly depend 
on the interchain coupling, as shown in Fig.\ \ref{fig:phase-2}.

Moreover the value of the Haldane gap is much smaller in the $S=2$ spin chain 
than for $S=1$. This is known reliably for the isotropic antiferromagnetic 
model only\cite{white,schollwock}, but should be true in the anisotropic 
case as well. We show schematically in Fig.\ \ref{fig:gaps}, how the 
antiferromagnetic and Haldane gaps depend on the anisotropy for the 
spin-1/2, spin-1 and spin-2 Heisenberg models. In all cases the 
antiferromagnetic state has a doubly degenerate ground state with a 
finite energy gap to the spin wave excitations. For $S=1/2$ the 
transition into this state is of Kosterlitz-Thouless type, and the gap
opens exponentially slowly. For spin-1 and spin-2 chains the transition 
from the Haldane phase is of second order, so the gap opens linearly on
both sides. The vanishing of the Haldane gap at $(J_z/J_{xy})_{c2}$ 
happens again in a Kosterlitz-Thouless-like manner.

By coupling now two spin-1 chains into a spin-1 ladder, the effect of the
interchain coupling is expected to be opposite to that discussed above. 
The Haldane gap decreases or it might even vanish when this coupling is 
switched on. This raises the question whether in the spin-1/2 ladder model
the interchain coupling is 
in fact a relevant perturbation for $-1 \leq J_z/J_{xy}<0$, and perhaps 
the phase boundary between
the Haldane and planar phases does not go 
in the way discussed above. An alternative possibility is shown in 
Fig.\ \ref{fig:phase} by a  dashed line.

\section{Numerical Procedure}

In this paper we try to elucidate this problem by looking at the 
degeneracy of the ground state and the generation or disappearance of 
gaps when the interchain couplings are switched on. In order to do this 
we have applied the DMRG method \cite{white} to the model defined by 
Eqs.\/ (\ref{eq:h0}) and (\ref{eq:h1}) using the anisotropy
$J_{z}/ J_{xy}$ and the relative strength of the interchain coupling, 
$\lambda$ as the two parameters. 

It has been found earlier in various applications, that the DMRG method gives
better results for systems with open boundary condition (OBC) than for
periodic boundary condition (PBC). On the other hand, when OBC is used,  
the free spins remaining at the ends in the valence-bond like state of the 
ladder model introduce extra energy-level degeneracies, and therefore
the analysis of the spectrum becomes somewhat more difficult.
In that region of the couplings, where the gap is very small, in order to get 
more accurate results, we have used OBC, otherwise we have done 
calculations with both boundary conditions.

The major limiting factor in the application of the DMRG method is that 
only a subset of all possible states is kept, and the eigenvalue matrix is
diagonalized in this subspace only. The accuracy can be improved, if first
the Hilbert space is block diagonalized by taking into account the
natural symmetries of the Hamiltonian, and the DMRG procedure is applied
to the blocks separately, considering the same fixed number of states in 
each block. When using OBC, translational symmetry is lost, so the total 
momentum is not a good quantum number. In the anisotropic model we also 
loose the $SU(2)$ symmetry. Therefore only the $z$ component of the total 
spin, $S_T^z$ can be used to classify the energy levels. 

It is easily seen that the ladder model Hamiltonian has two further 
symmetries, two mirror planes that are perpendicular to the plane of the 
ladder. One goes across the middle of the rungs of the ladder. It 
transforms the $\sigma$ and $\tau$ spins into each other ($\sigma$--$\tau$
symmetry). The other cuts the legs of the ladder in the middle, reflecting
the spins on the left onto spins on the right (left--right symmetry). 
Moreover, the $S^z_T=0$ subspace, which will be of particular interest for
us, has an additional symmetry, the spin reversal or up--down symmetry. 
The corresponding operators will be denoted as ${\cal P}_{\sigma-\tau}$, 
${\cal P}_{l-r}$, and ${\cal P}_{\uparrow \downarrow}$. The eigenstates 
of the system can be either symmetric or antisymmetric under these operations.
An even parity state will be labelled by $+$ while an odd parity state
by $-$. 

It is advantageous to include these symmetries in the DMRG procedure
by choosing wave functions that are eigenstates of the parity operators. 
Using the notation of White \cite{white4}, the
superblock wave function $\psi (\alpha_l s_{l+1} s_{l+2} \beta_{l+3})$  
is formed out of the states $\alpha_l$ and $\beta_{l+3}$  of the two blocks 
and the states $s_{l+1}$ and $s_{l+2}$ of the two new added spins, 
\begin{eqnarray}
     &\mid& \psi_{\pm} \rangle = 
     \sum_{\alpha_l s_{l+1} \atop s_{l+2} \beta_{l+3}} 
   A_{\alpha_l s_{l+1} s_{l+2} \beta_{l+3}}  \nonumber \\ 
   && \times  \left\{ \psi (\alpha_l s_{l+1} s_{l+2} \beta_{l+3}) \pm 
    {\cal P}  \psi (\alpha_l s_{l+1} s_{l+2} \beta_{l+3}) \right\} \/.
\label{eq:parity}
\end{eqnarray}
$A_{\alpha_l s_{l+1} s_{l+2} \beta_{l+3}}$ is a normalization 
constant which takes the value $1/2$ or $1/\sqrt{2}$, depending on 
whether the state obtained by the symmetry operation is identical 
with the original one or not.

The $\sigma$--$\tau$ and spin reversal operations are products of local
operators that act on the spin states of a single rung, so they can
be used even if the left and right blocks are not of equal length.
The four states $|\! \downarrow \downarrow \rangle$,
$|\! \downarrow \uparrow \rangle$, $|\! \uparrow \downarrow \rangle$ 
and  $|\! \uparrow \uparrow \rangle$ of the $\sigma$ and $\tau$ spins on 
the same rung span a basis on which 
\begin{equation}
 {\cal P}_{\sigma - \tau}^1 = \left( \matrix{  
    1 & 0 & 0 & 0  \cr
    0 & 0 & 1 & 0  \cr
           0 & 1 & 0 & 0  \cr
    0 & 0 & 0 & 1 } \right) \/, \quad 
 {\cal P}_{\uparrow \downarrow}^1 = \left( \matrix{  
    0 & 0 & 0 & 1  \cr
    0 & 0 & 1 & 0  \cr
           0 & 1 & 0 & 0  \cr
    1 & 0 & 0 & 0 } \right) \,.
\end{equation}

In each iteration cycle of building larger blocks the operators are 
renormalized  as $P^{l+1}=O P^{l} O^{\dagger}$ and they are stored with 
the other block operators. The $O$ matrix contains the new truncated 
basis states of the subsequent iteration cycles. 
A local operator acting on a block state gives
\begin{equation}
    P^l | \alpha_l \rangle =  a_{\alpha_l}| \alpha_l' \rangle  \,,
\end{equation}
where the coefficient $a_{\alpha_l}$ is $1$ or $-1$ for the symmetric or 
antisymmetric combination, respectively. 

In the first step of the infinite algorithm, when the block
consists of the states of a single rung, 
$| \alpha_1 \rangle =|s_{l+1}\rangle $ and $a_{\alpha_l}=1$.
The iterated matrix $P^{l+1}$ has one element in each row and that  
value gives $a_{l+1}$. Therefore, the local symmetries simply give 
\begin{eqnarray}
   \lefteqn{  {\cal P}_{local}^l  \psi (\alpha_l s_{l+1} s_{l+2} 
     \beta_{l+3}) =   } \hspace{1cm} \nonumber \\
    & &  a_{\alpha_l} a_{\beta_{l+3}}
  \psi (\alpha_l' s_{l+1}' s_{l+2}' \beta_{l+3}') \/.  
\end{eqnarray}

On the other hand the operator ${\cal P}_{l-r}^l$ mixes the states of 
different rungs and can be used only when the two blocks are of equal
length. When acting on a superblock wave function, it gives
\begin{equation}
{\cal P}_{l-r}^l  \psi (\alpha_l s_{l+1} s_{l+2} \beta_{l+3}) = 
     \psi (\beta_{l+3} s_{l+2} s_{l+1} \alpha_{l})  \/.
\label{eq:Plr}
\end{equation}

The total parity of the wave function is the product of the left--right, 
$\uparrow \downarrow$ and $\sigma$--$\tau$ parities. In what follows, 
unless otherwise specified, only the total parity will be given. It is 
worth mentioning that if several symmetries are used, the factor 
$A_{\alpha_l s_{l+1} s_{l+2} \beta_{l+3}}$ has to be included for each 
symmetry. This ensures that every configuration is taken into account 
only once.

All these symmetry operations reduce the size of the Hilbert space of the
superblock Hamiltonian by a factor of about two. This decreases the memory
requirements and allows to keep more spin configurations, which improves the 
accuracy of the DMRG method. 
The local symmetries can be included into both the infinite- and 
finite-system algorithms. The left-right symmetry can be used in the 
infinite-system method only, but by keeping more states it would
produce better starting vectors \cite{white4} for the diagonalizaton 
procedure in the finite-system method. 

Using these symmetry considerations the Hilbert space is split into
subspaces characterized by the $z$ component of the total spin and the
parity. In order to obtain the ground state and some low lying states of 
the ladder, in principle one has to determine a few low lying eigenstates 
of all subspaces. For a finite chain with $N$ sites the $n^{th}$ energy 
level of the $S_T^z=s$ sector with parity $p$ will be denoted by 
$E_{s,n,p}(N)$. Similarly the energy gap between the states with energies 
$E_{s',n',p'}$ and $E_{s,n,p}$ will be denoted by 
$\Delta_{s,n,p;s',n',p'}(N)$.

Except for the ferromagnetic regime, in the spin-1/2 chain the lowest 
lying energy level belongs to the $S_T^z=0$ sector and it has odd parity 
under left-right and spin reflection symmetries for $N \bmod 4=0$ while 
it has even parity for chains with $N \bmod 4=2$. The parity of the first 
excited state is opposite to that of the ground state and the parity 
changes for every higher lying level. Therefore, in the infinite-lattice 
algorithm of the DMRG method the parity has to be changed in every 
iteration cycle.

This difficulty does not arise either in the spin-1 chain or in the ladder
models. In the spin-1 chain the ground state was found to be in the $S_T^z=0$ 
sector with even parity under left-right reflection for all chain lengths 
$N$. In the ladder model beside the left to right reflection we also 
have the $\sigma$--$\tau$ symmetry. The ground state of this model has 
positive parity under both reflections for all non-negative values 
of $\lambda$. The excited states also preserves their parity as a 
function of $\lambda$. Therefore in this case the same symmetry combination 
has to be used in every successive step of the DMRG algorithm.     

Because the energy gaps are expected to be rather small close to the phase
boundaries, we have used the finite-lattice method version
of DMRG with two or three iteration cycles. We calculated the energy
gaps for ladders with $N=16, 32, 48, 64, 82, 100, 128$ sites,
and used a finite-size scaling procedure to extrapolate to infinite 
system. Due to our restricted computational facility the number
of states $M$ representing the block in the DMRG method could
be chosen between $M=100$ and $200$ states. The truncation error
was worst close to the critical points. For the second and 
higher excited states it was about $10^{-6}- 10^{-7}$, corresponding
to a real error of about $10^{-3}$.

Several formulae have been proposed to extrapolate the finite-size 
results to the thermodynamic limit. When a finite gap is expected, 
the $N\rightarrow\infty$ limit of the gap can be obtained by fitting 
the energy difference of the various $E_{s,n,p}$ energy levels to
the form \cite{hida} $\Delta (N)= \Delta + A \exp (- N/ \xi )$. 
On the other hand, for open systems, where the finite size effects are
more pronounced, the form\cite{schollwock} $\Delta(N)=\Delta+A/N$
or an inverse squared dependence\cite{sorensen} $\Delta(N)=\Delta+A/N^2$
should be used. The former one is known to give a lower-bound estimate of 
the gap. 

In most cases we have used the formulae with $1/N$ or $1/N^2$. The 
exponential dependence was assumed when neither the $N^{-1}$ nor the 
$N^{-2}$ fit gave satisfactory result.

\section{Numerical Results}

In this section we present the results of our numerical calculations. 
Since the gaps are expected to be rather small, resulting in relatively
large errors in the extrapolation to infinite systems, we are not able 
to locate exactly the phase boundaries, where the gaps vanish. We will 
therefore restrict ourselves to a qualitative check of how the phase 
diagram may look like.

Because nothing particularly interesting is expected on the $J_z/J_{xy}>0$
side, the system is ferromagnetic if $J_z/J_{xy}>1$ and planar if
$0 < J_z/J_{xy} < 1$ independently of $\lambda$, in what follows we will
consider the $J_z/J_{xy}< 0 $ region only. We have chosen four 
different values of the anisotropy, namely $J_z/J_{xy}=-2, -1.1, -0.8, -0.5$ 
and varied $\lambda$ from 0 to 1, to see how the character of the 
ground state changes. The energies will always be measured in units of
$J_{xy}$.

\subsection{The $\lambda=0$ and $\lambda=1$ lines}

As mentioned above, at $\lambda=0$ the properties of two decoupled 
spin-1/2 chains should be recovered, while at $\lambda = 1$ at least
the low energy part of the spectrum is identical to that of the spin-1
chain. Our calculations along these lines aimed to test the accuracy
of the method and to find the relevant subspaces of the Hilbert space.

Let us consider first the case $\lambda=0$. When $J_z/J_{xy}<-1$,
the antiferromagnetic chains give a $2\times 2$-fold degenerate ground
state in the $S_T^z =0$ subspace. Accordingly the gap $\Delta_{02-,01+}$ and
two other gaps in the $S^z_{T}=0$ sector should go to zero in the limit 
$N\rightarrow \infty$. In fact our calculation gives an extrapolated 
value of the order of $10^{-3}$. This shows the real accuracy of the 
numerical calculations. 

The next levels above this fourfold degenerate ground state have one 
or two spins flipped. The corresponding gaps, $\Delta_{11-,01+}$  and 
$\Delta_{21+,01+}$ were found to be finite, opening slowly as the
anisotropy increases. Above this gap a continuum of states is found 
as shown by the vanishing of the gap $\Delta_{12+,11-}$.

In the planar phase all these gaps should vanish. In fact in the range
$ -1 < J_z/J_{xy} < 0 $ the gaps $\Delta_{11-,01+}$ and $\Delta_{21+,01+}$ 
were found to be of the order of $10^{-3}$, which we consider to be zero.

Along the line $\lambda=1$ the antiferromagnetic phase appears for
$J_z/J_{xy} < -1.18$. The twofold degeneracy of the ground state is in 
fact recovered, since the gap $\Delta_{02-,01+}$ scales to values
of the order of $10^{-3}$. The antiferromagnetic gap $\Delta_{11-,01+}$ 
remains finite. 

In the Haldane phase, $-1.18 < J_z/J_{xy} < 0$, we have again a fourfold 
degenerate ground state, if the ladder has free ends, but here the 
gaps $\Delta_{02-,01+}$ and $\Delta_{11-,01+}$ should scale to zero 
exponentially. The values obtained for the gap are of the order of 
$10^{-4}$. The next levels above them are $E_{21+}$ and $E_{12+}$, which 
cross each other at $J_z/J_{xy}=-1$. Therefore the real gap above the
ground state is $\Delta_{21+,01+}$ for $J_z/J_{xy}>-1$,  while 
it is $\Delta_{12+,01+}$ for $J_z/J_{xy}<-1$. This gap was found to be 
0.147(1), 0.4105(3), 0.223(1), 0.096(3) for $J_z/J_{xy}=-1.1, -1, -0.8, -0.5$. 
Our result at $J_z/J_{xy}= -1$ agrees quite well with the best estimate
\cite{white} for the Haldane gap of the isotropic $S=1$ model. 

\subsection{The antiferromagnetic regime}

Taking now intermediate values for $\lambda$, we consider first  the 
transition from the Haldane phase to the antiferromagnetic phase for 
$J_z/J_{xy}<-1$. Since the phase boundary is not determined by any symmetry, 
it is expected that for $-1.18<J_z/J_{xy}<-1$ the antiferromagnetic gap,
$\Delta_{11-,01+}$ which is finite at $\lambda = 0$, should vanish for a 
finite $\lambda_{c}$, and then for larger $\lambda$ the Haldane gap 
$\Delta_{12+,01+}$ should open up. Since the values of the gaps are very 
small, it is difficult to determine accurately the value $\lambda_{c}$, 
where this happens. 

As a first attempt, we have investigated the closing of the 
antiferromagnetic and the opening of the Haldane gap as a function of
$\lambda$ at $J_z/J_{xy} = -1.1$. The thermodynamic limit of the Haldane 
gap could be determined very well using the $N^{-2}$ dependence for 
$\lambda > 0.1$. In this region this gap is finite. At the same time
the antiferromagnetic gap is found to vanish. For $\lambda \leq 0.1$, however, 
no reliable extrapolation procedure was found and therefore we can only 
claim that the transition from the antiferromagnetic to the Haldane phase 
occurs at a very small value of $\lambda$, between $\lambda=0$ and 
$\lambda=0.05$. This indicates, that due to the smallness of the 
antiferromagnetic gap of the spin-1/2 chain, a very small although 
finite interchain coupling is sufficient to drive the system into the 
Haldane phase. This means that the phase boundary starts from the 
$J_z/J_{xy}= -1$, $\lambda=0$ point with an almost horizontal slope.
 
Taking a somewhat larger anisotropy, $J_z/J_{xy}=-1.15$, the
antiferromagnetic state survives to a larger value of $\lambda$, 
as shown in  Fig.\ \ref{fig:d1_15}. The lower bound estimate of the gap 
$\Delta_{11-,01+}$ at $\lambda=0.1$ is $0.030(3)$. This point, therefore, 
belongs to the antiferromagnetic phase. Moving up in the phase diagram 
with $\lambda$ to $J_z/J_{xy}=-1.15$, $\lambda=1$ or to the right to the
point $J_z/J_{xy}=-1.1$, $\lambda=0.1$, this same antiferromagnetic gap 
clearly scales to zero, indicating that these points are already in 
the Haldane phase.  The phase boundary should therefore have the shape 
indicated in Fig.\ \ref{fig:phase}.

\subsection{The planar phase}

Next we considered the stability of the planar phase in the 
$-1 < J_z/J_{xy} < 0$ regime. We have measured the gap $\Delta_{21+,01+}$, 
which should vanish at $\lambda=0$, as a function of $\lambda$ for several 
values of the anisotropy. 

The accuracy of our calculations was checked by comparing the energy values 
calculated for $\lambda < 1$ and $\lambda > 1$, using the self-duality 
relationship of Eq.\ (\ref{eq:self}). We have got the same energies
to at least 6 or 8 digits. We show in Fig.\ \ref{fig:d0_8} the results 
obtained for $J_z/J_{xy}=-0.8$. At $\lambda = 0 $ the gap scales to 
$0.008(2)$ as $1/N$. At $\lambda=0.2$ an excellent fit can be obtained by the 
$1/N^2$ fit for $N=128,100,82$. The extrapolated gap was found to be 
$0.057(2)$. 

The finite-size scaling procedure is not that clear for $\lambda=0.1$.
Neither a $1/N$ nor a $1/N^2$ fit is satisfactory. This indicates that the
available chain lengths are still not long enough to determine 
$\Delta$ accurately. The estimated value of the gap is 
between $0.025(3)$ and $0.035(3)$. Comparing this with the gap at
$\lambda=0.2$, we could conclude that in the thermodynamic limit the 
gap opens linearly with $\lambda_c=0$, in the same way\cite{legeza} as
for $J_z/J_{xy}=-1$.

For $J_z/J_{xy}$ closer to zero, the Haldane gap is too small even at 
$\lambda=1$. At $J_z/J_{xy}=-0.5$ the extrapolated value of the gap
is about $0.096(3)$. For small values of $\lambda$ of the order of $0.1$ 
the error of the DMRG method becomes comparable to the value of the gap 
and no reliable extrapolation procedure was found. The results are, however,
in agreement with the assumption that $\lambda_c = 0$. 

These results indicate that the interchain coupling is relevant in the 
whole $-1 \leq J_z/J_{xy} < 0$ region. An arbitrarily weak interchain 
coupling of this type would drive the spin-1/2 ladder into the Haldane phase.

\subsection{The spin-1 ladder model}

The spin-1 ladder model is constructed in the same way as the spin-1/2 
model except that the $\sigma_i$ and $\tau_i$ operators in Eq. (\ref{eq:h0}) 
and (\ref{eq:h1}) are now spin-1 operators. In the $\lambda=0$ limit, 
which in this case corresponds to two decoupled spin-1 chains, the 
Haldane phase exists for a wide range of anisotropy, namely for 
$-1.18<J_z/J_{xy}<0$. At $\lambda = 1$ this model should behave like 
a spin-2 chain, in which the Haldane phase is squeezed into a very narrow 
region around the isotropic point. The gap itself is also small. 
According to the best estimate\cite{schollwock} it is 
$\Delta=0.085(5)$ at the isotropic point. 

Therefore, as shown if Fig.\ \ref{fig:phase-2}, the Haldane gap is
expected to vanish at a finite value of the interchain coupling when 
$J_z/J_{xy}$ is not too close to $-1$. In order to see this,
we have looked at the gap $\Delta_{21+,01+}$, which is finite in the
Haldane phase, but vanishes in the planar one. It was found that 
at $J_z/J_{xy}=-0.7, \lambda=0.5$ this gap was already zero.
Although we could not locate the boundary precisely, it is clear that,
taking into account again the self-duality relationship, the phase 
boundary should be perpendicular to the anisotropy axis at $\lambda=1$ and
have the shape shown in Fig.\ \ref{fig:phase-2}. 

\section{Discussion}

In this paper we have studied the stability of the Haldane phase in a 
magnetic ladder model. Our special choice of the coupling between the 
chains allowed us to investigate and smoothly interpolate between both 
the half-odd integer and integer spin Heisenberg models. We have applied 
the DMRG method which was improved to reduce the computational time and to 
increase the accuracy.

We have found that for the $S=1/2+1/2$ composite spin model 
the phase boundary between the antiferromagnetic and the Haldane phase
is a smooth curve and the value of $\lambda_c$ is a function of the 
anisotropy parameter, $J_z/J_{xy}$. For the boundary between 
the Haldane and planar phases our results indicate that it is at
$J_z/J_{xy} = 0 $ independently of $\lambda$. For $0 \leq J_z/J_{xy} < 1$
the system remains in the planar phase, while for $-1 \leq J_z/J_{xy}<0$
the interchain coupling is relevant. The critical value where the Haldane 
gap starts to open up is $\lambda_c = 0$. 

This is, however, not the only possibility. Our results are not in 
contradiction with the assumption that $\lambda_c$ is finite,
although small, except when $J_z/J_{xy}$ is rather close to $0$, as
shown by the dashed line in Fig.\ \ref{fig:phase}.

 We should also mention 
that since the point $\lambda=1, J_z/J_{xy}=0$ on the boundary between the
Haldane and planar phases was obtained in the continuum limit, it cannot
be excluded that in the lattice model this boundary at $\lambda=1$ is
not exactly at $J_z/J_{xy}=0$, but at a small negative value.
In this case the boundary between the Haldane and planar phases should
be slightly deformed from that shown in Fig.\ \ref{fig:phase}.



On the basis of our calculations, unfortunately, we are unable to 
decide, which scenario takes place. The argument that gave 
$(J_z/J_{xy})_{c2}= 0$ at $\lambda=1$ would give a phase boundary
which is independent of $\lambda$, meaning that the interchain coupling
is relevant for any $J_z/J_{xy}$ in the range $ -1 \leq J_z/J_{xy} < 0 $.
The same reasoning might then indicate that the interchain coupling 
is relevant also in the case when four $S=1/2$ chains are coupled. With an 
appropriate choice of the couplings this would, however, generate a 
four-leg ladder that behaves like an $S=2$ chain. In this model
the phase boundary between the planar and Haldane phases cannot be
independent of $\lambda$, in the same way as in the two-leg $S=1$ ladder. 
Further calculations, keeping probably close to 1000 states in the DMRG 
procedure, are necessary to decide between these possible
scenarios and to determine the precise shape of the phase boundaries.

\section{Acknowledgments}

This research was supported in part by the Hungarian Research Fund 
(OTKA) under Grant No.\ 15870 and by the U.S.-Hungarian Joint Fund
under Grant No.\ 555.

\newpage


\begin{figure}
\caption{Schematic phase diagram of the $S=1$ ladder model in the
$(\lambda, J_z/J_{xy})$ plane.}
\label{fig:phase-2}
\end{figure}

\begin{figure}
\caption{The energy gap of the $S=1/2, 1$ and 2 Heisenberg chains
as a function of the anisotropy $J_z/J_{xy}$.}
\label{fig:gaps}
\end{figure}

\begin{figure}
\caption{The antiferromagnetic gap $\Delta_{11-,01+}$ as a function of 
 $N^{-1}$ at (a) $J_z/J_{xy}=-1.15$, $\lambda =0.1$ and $\lambda=1$ and (b)
$J_z/J_{xy}=-1.1$, $\lambda=0.05$ and $\lambda =0.1$.}
\label{fig:d1_15}
\end{figure}

\begin{figure}
\caption{The energy gap $\Delta_{21+,01+}$ as a function of
 $N^{-2}$ at $\lambda=0.2$, $J_z/J_{xy}=-0.8$.}
\label{fig:d0_8}
\end{figure}

\end{document}